\documentclass[12pt,twoside]{article}
\usepackage{fleqn,espcrc1}

\usepackage{graphicx}
\usepackage[figuresright]{rotating}

\newcommand{\AmS}{{\protect\the\textfont2
  A\kern-.1667em\lower.5ex\hbox{M}\kern-.125emS}}

\title{Instrumentation for and First Results on Nuclear Responses 
       for Supernova Explosions}

\author{H.J. W\"ortche\address{Institut f\"ur Kernphysik, 
                 Westf\"alische Wilhelms-Universit\"at,\\ 
                 Wilhelm-Klemm-Strasse 9, 48149 M\"unster, Germany}
                 \thanks{Present address: Kernfysisch Versneller Instituut, 
                         Rijksuniversiteit Groningen, Groningen, 
                         The Netherlands, E-mail : wortche@kvi.nl}\\
                 for the EUROSUPERNOVA collaboration$^{**}$}
\begin{document}

\maketitle

\begin{abstract}
Our collaboration has set up a focal plane detection system and a 
focal plane polarimeter at the large acceptance Big-Bite Spectrometer
at AGOR. The detector systems are equipped with a high performance 
readout and online data processing system, which allows polarization 
transfer and charge transfer measurements at extreme forward angles
with high precision. Preliminary results on GT$_+$ strength distributions 
obtained in $(d,^2\!\mbox{He})$ measurements revealing the fine structure 
of the distributions are presented. Their relation
to recent calculations of stellar weak interaction rates is discussed.    
\end{abstract}

\section{Introduction}
In the past years great efforts have been made to improve the 
reliability of theoretical predictions used to calculate stellar weak 
interaction rates. Special emphasis has been put on allowed and 
first-forbidden Gamow-Teller (GT) and Fermi (F) transitions, 
which govern electron-capture, positron-capture and $\beta$-decay 
rates, as well as $\nu$-nucleosynthesis in different stages of a 
supernova event. To obtain predictions with significantly improved 
reliability, as compared to the standard work performed by Fuller,
Fowler and Newman \cite{fuller80,fuller82,fuller82a,fuller85},
rates have recently been determined based on systematic large-scale 
shell model calculations in the mass range A = 45-65
\cite{caurier99,langanke00}. The reliability of the calculations 
has been tested by comparing the excitation spectra
as well as the GT$_-$ and GT$_+$ strength distributions 
with experimental data. In case of the GT strengths, 
the crucial condition was a proper reproduction of the GT strength 
distributions deduced from $(p,n)$ and $(n,p)$ measurements and 
emphasis was put on transitions leading to the population of  
low-lying states, where due to phase space enhancement the (fine) structure 
of the distributions gets relevant \cite{caurier99}.  
\par
With this theoretical tool available, experiments are
needed, which provide information on GT strength distributions 
with improved precision, compared to the pioneering $(p,n)$ and
$(n,p)$ experiments (for the nuclei of consideration see 
\cite{caurier99} and references therein). 
A successful experimental approach has been
demonstrated in Refs. \cite{fujita96,fujita97}. A combined
analysis of high resolution $(^3\mbox{He}, t)$, $(p,p')$ and  
$(e,e')$ data yielded information on the structure 
of the GT strength distribution and the separation of isospin 
components in $^{28}$Si and $^{58}$Cu. 
\par   
Following the approach of combining the spin-isospin selectivity of 
hadronic probes, the EUROSUPERNOVA collaboration
has set up and commissioned a focal plane detection system 
(FPDS) and a high performance focal plane polarimeter (FPP) 
at the large acceptance Big-Bite Spectrometer (BBS) \cite{berg95} at 
AGOR. Taking advantage of various particle beams provided at medium energies
($E/A \simeq 100 - 200$ MeV) by the AGOR cyclotron, the system 
is bound to perform polarization transfer measurements in 
inelastic proton scattering and charge exchange reactions like 
$(d, ^2\mbox{He})$ and $(^3\mbox{He},t)$. The experiments are performed
at extreme forward angles, including $0^\circ$, where  
due to the momentum dependence of the hadronic interaction 
the scattering becomes especially sensitive for excitation of  
spin-flip transitions.
\par
In section \ref{sec:setup} of this contribution, we present a short 
overview of the detector setup, the detector readout and the 
Digital Signal Processor (DSP) based data acquisition system.
In section \ref{sec:exp_res}, preliminary experimental results are
presented. For details of the setup we refer to 
Refs. \cite{kruesemann99,hagemann99,kruesemann00}.   
\section{The EUROSUPERNOVA detector system} 
\label{sec:setup}
\subsection{Detector setup}
\begin{figure}[tb]
	\begin{center}
		\resizebox{10.5cm}{!}
		 {\includegraphics{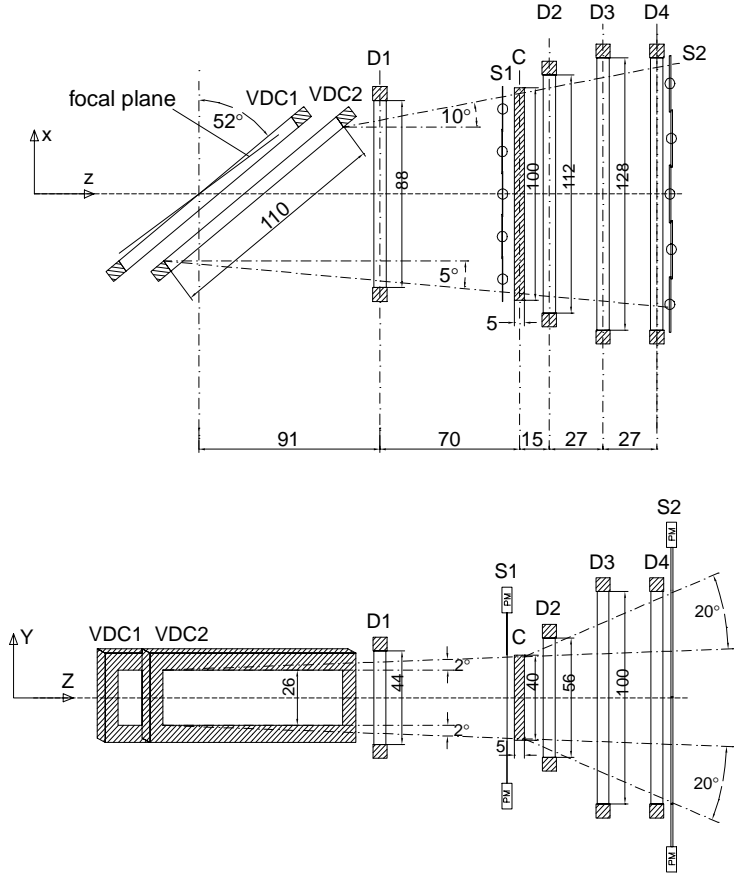}}
	\end{center}

\caption{Top view (upper panel) and side view (lower panel) of the 
         EUROSUPERNOVA detector. The FPDS 
         consists of two VDC's, tilted $52^\circ$ to the normal of the
         central beam. Tracking in the FPP is performed by four MWPC's
         D1 - D4. Two segmented scintillator arrays are labelled as S1 
         and S2, the graphite analyzer is labelled as C. Lengths are 
         indicated in units of centimeters. (Figure adapted from Ref. 
         \cite{hagemann99}).}   
\label{fig:fpp_scheme}
\end{figure}

In Fig.\ref{fig:fpp_scheme} a schematic layout of the 
EUROSUPERNOVA detector is depicted. In order to deduce the 
momentum vector, scattered particles are detected near the
BBS focal plane using a set of Vertical Drift Chambers (VDC's) VDC1
and VDC2. The momentum acceptance of the system with the BBS 
in mode B is about 15\%. The angular acceptance 
is 66 mrad in the horizontal and 140 mrad in the vertical direction. 
The energy resolution achieved for 150 MeV protons is 
100 keV FWHM. 
\par

If the detector is operated in polarimeter mode, the 
polarization of scattered particles is deduced by measuring the 
asymmetry of 2$^{nd}$ scattering in the graphite analyzer C. 
Tracking upstream and downstream of the analyzer is performed by 
virtue of a set of Multi Wire Proportional Chambers (MWPC's) D1 - D4. 
The angular acceptance for 2$^{nd}$ scattering is limited to 
angles smaller than $20^\circ$. 
\subsection{Detector readout}
\begin{figure}[tb]
	\begin{center}
		\resizebox{\textwidth}{!}
		 {\includegraphics{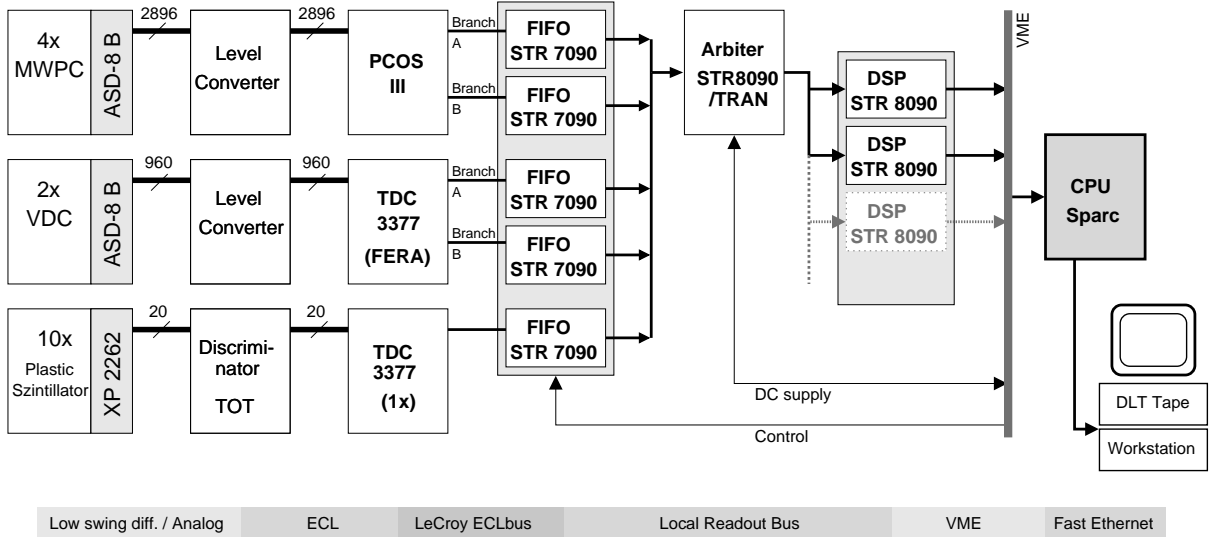}}
	\end{center}

\caption{Scheme of the DSP-based data acquisition system. The amount 
         of active channels, sense wires for wire chambers, 
         photomultipliers for scintillator arrays, are indicated.  
         (Figure adapted from Ref. \cite{hagemann99}).}  
\label{fig:fpp_dsp}
\end{figure}

Operation at extreme forward angles causes high rate loads in 
the detection system. In addition, the rates in different sections 
of the wire chambers may differ as much as 5 orders of magnitude.
To guarantee a flat efficiency, the wire chambers have been 
equipped with newly developed ASD-8 based preamplifiers, which are 
capable of processing data up to 30 MHz without significant 
pile-up \cite{kruesemann00}. The preamplifiers provide differential 
low-swing ECL signals, which are transferred via ECL level converters 
to the CAMAC based front-end electronics (see Fig.\ref{fig:fpp_dsp}).
The integrated charge information is maintained in the width of 
the signals, a feature which might allow energy-resolving detection 
in future applications. The MWPC data are processed by 
the LeCroy PCOS III system, the VDC data by the LeCroy pipeline
TDC system 3377 \cite{lecroy}. Readout of the complete detector,
in total about 4000 wires, is accomplished in less then 7 $\mu$s 
for a standard event.     
\subsection{Data acquisition and online processing}
In inelastic proton scattering, a major constraint performing 
polarization transfer measurements is the by far dominant small-angle Coulomb 
scattering in the analyzer. This effect limits the angular range, 
where usable asymmetries can be obtained to angles larger than $5^\circ$ 
and reduces the polarimeter efficiency to about 5 \%. 
\par
In the present setup, a DSP based data acquisition system has been implemented,
which performs a readout of {\sl each} event triggered by a coincidence 
of scintillator planes S1 and S2 and subsequently performs a software
evaluation of the data. This solution offers great flexibility. Changing
the detector e.g. from $(\vec{p}, \vec{p}\,')-$mode 
to $(d,^2\!\mbox{He})$-mode,
described below, solely requires reprogramming of the DSP's and removal of the 
graphite analyzer. Also operating the detector 
in coincidence to other detector systems is greatly simplified
and can be achieved through synchronization to the 1$^{st}$ level trigger,
as has been successfully tested for the 
$^{26}$Mg($^3$He, t$\gamma$)$^{26}$Al$^*$ reaction \cite{sakoda99}.
\par
Maximum performance is achieved by matching the processing and 
the readout time and decoupling the DSP's from the random front-end
data by a FIFO system. The present setup is capable of handling an 
incoming rate up to several hundred kHz, with the DSP not contributing 
to the system dead time \cite{hagemann99}.  
\section{Preliminary experimental results}
\label{sec:exp_res}
After commissioning the setup, our collaboration has performed 
high statistics cross section and polarization transfer measurements of 
inelastic proton scattering from $^{11}$B, $^{12}$C, $^{48}$Ca, $^{58}$Ni and 
$^{124}$Sn. The experiments are mainly geared to extract spin-flip M1 
and spin-dipole strengths distributions up to 30 MeV excitation, the
data are presently analyzed.
\par
In addition, we have started to investigate the GT$_+$ strength 
distribution in a A=45-65 mass nucleus via the reaction
$^{58}$Ni$(d,^2\!\mbox{He})^{58}$Co and supplementary calibration 
measurements  $^{12}$C$(d,^2\!\mbox{He})^{12}$B  and 
$^{24}$Mg$(d,^2\!\mbox{He})^{24}$Na at scattering angles 
$\theta=0^\circ-27^\circ$. 
\par
The performance of $(d,^2\!\mbox{He})$ measurements at extreme forward
angles including $0^\circ$ is a non trivial exercise, because
it requires coincident detection of correlated proton pairs
originating from the unbound $^2$He-system in the vicinity of a 
dominant background caused by deuteron breakup protons.
\begin{figure}[ttt]
	\begin{center}
		\resizebox{\textwidth}{!}
                {\rotatebox{-90}
		 {\includegraphics{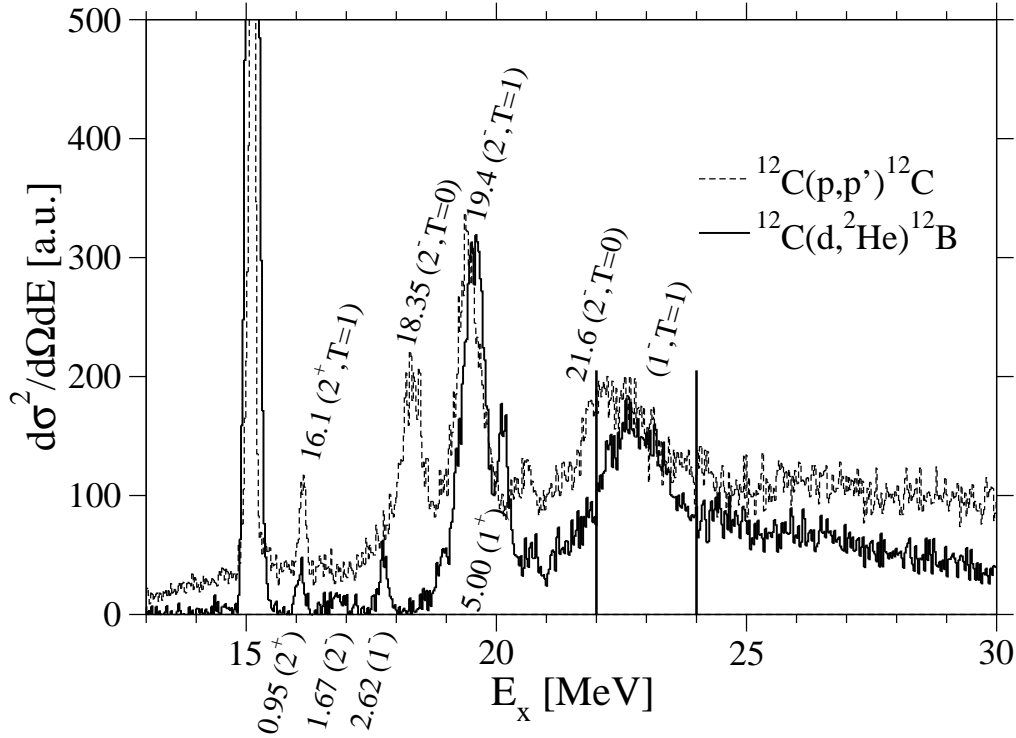}}}
	\end{center}
\vspace*{-20mm}
\caption{Double differential $^{12}$C$(p,p')^{12}$C and  
         $^{12}$C$(d,^2\!\mbox{He})^{12}$B cross section spectra  
         measured for E$_p$=172 MeV, $\theta = 10^\circ$ and 
         for E$_d$=170 MeV, $\theta = 0^\circ$, respectively.  
         The $(d,^2\!\mbox{He})$ cross sections are not acceptance corrected 
         and are shifted by 15.11 MeV (see text). 
         Prominent excitations in $^{12}$C \cite{ajzenberg90} are indicated
         above, prominent excitations in $^{12}$B \cite{ajzenberg90}
         below the spectra.
         The latter excitation energies refer to the $^{12}$B ground state.
         The vertical lines mark a range of $^{12}$C $(J^\pi,T=1^-,1)$ 
         excitations.} 
\label{fig:12c}
\end{figure}
In order to operate the EUROSUPERNOVA  detector in  $(d,^2\!\mbox{He})$-mode,
the graphite analyzer is removed and the DSP's are reprogrammed to identify
tracks of correlated protons by means of the MWPC information. 
Due to the limited angular acceptance of the BBS, detected $^2$He 
proton pairs originate from the target with small (relative) kinetic
energy, i.e. the $^2$He system is nearly exclusively detected
in $^1S_0$ configuration. The $(d,^2\!\mbox{He})$ reaction therefore acts as a 
filter for $\Delta S = \Delta T =1$ transitions,
if a simple one-step reaction mechanism is assumed.
\par
Figure \ref{fig:12c} shows a $^{12}$C$(d,^2\!\mbox{He})^{12}$B 
double-differential cross section spectrum measured at 
$0^\circ$ and E$_d$=170 MeV
with a $(p,p')$ double-differential cross section spectrum measured 
at  $10^\circ$ and E$_p$=172 MeV overlaid. The data taking time for the
$(d,^2\!\mbox{He})$ spectrum amounted to 4 hours. 
The transition to the $^{12}$B ground-state and the analogue 15.11 MeV
spin-flip M1 transition in $^{12}$C have been matched in energy by 
shifting the $(d,^2\!\mbox{He})$ spectrum by 15.11 MeV.
\par
The energy resolution achieved for excitations in the residual nucleus
$^{12}$B is about 150 keV, revealing the fine structure
of the $^{12}$B spin-isospin response. The spectrum is free from background 
originating from random coincidence breakup protons. Both features could be 
achieved by applying a novel VDC analysis method based on imaging techniques
during offline analysis \cite{schmidt00}. 
The software is capable of identifying multiple
tracks in the VDC's based on the pipeline TDC event history, even in case of
spatial overlap, and provides the time difference of particles 
passing the VDC wire-planes. Based on the later information, the shape
of the random background is identified by correlating protons stemming
from different beam bursts in the analysis. 
\par 
The comparison illustrated in Fig.\ref{fig:12c} demonstrates, 
similar as in earlier measurements \cite{ohnuma93,xu95,inomata98}, 
the selectivity of the $(d,^2\!\mbox{He})$ probe for the 
excitation of spin-isospin modes. The energy resolution we obtain
nevertheless removes systematic uncertainties in identifying analogue
transitions, one of the major obstacles encountered in 
identifying GT$_+$ strengths distributions in the past.
\par
\begin{figure}[]
	\begin{center}
		\resizebox{\textwidth}{!}
                {\rotatebox{-90}
		 {\includegraphics{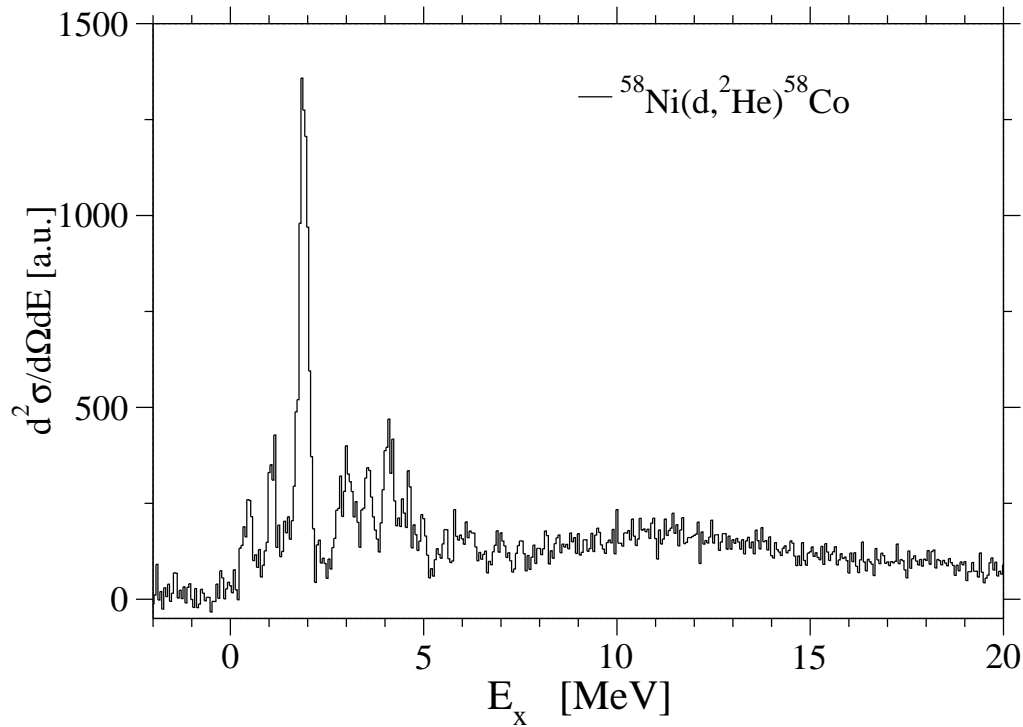}}}
	\end{center}
\vspace*{-10mm}
\caption{Double differential $^{58}$Ni$(d,^2\!\!He)^{58}$Co cross section
         spectrum. Excitation energies refer to the $^{58}$Co ground-state.
         The cross sections are acceptance corrected.}  
\label{fig:58ni}
\end{figure}
In Fig.\ref{fig:58ni}, we present a $0^\circ$ $(d,^2\!\mbox{He})$ 
cross section spectrum
measured at E$_d$=170 MeV for $^{58}$Ni, a nucleus in the 
mass range under consideration in Refs.\cite{caurier99,langanke00}. 
The spectrum is acceptance corrected, by modeling the 
BBS $^2$He acceptance in Monte-Carlo simulations. Up to 5 MeV excitation 
in $^{58}$Co, we detect a concentration of transitions, most prominent 
a strong transition at about 1.9 MeV. The transitions correlate to the 
broad distribution of GT$_+$ strength detected in $(n,p)$
measurements \cite{el-kateb94} and the analogue $T_0+1$ strength 
distribution deduced from $^{58}$Ni$(^3\mbox{He},t)^{58}$Cu data
\cite{fujita96}. Relating to the issue of the relative importance 
of low-lying transitions mentioned above, our data indicate distinct 
differences in the GT$_+$ strength distribution below 5 MeV compared to 
the earlier measurements, an issue which will be subject 
of further investigations and requires careful consideration of
the physics underlying the different probes.  
\par
At higher excitations the $(d,^2\!\mbox{He})$ cross section spectra reveal
the existence of a brought resonant structure centered at about
11 MeV, which becomes most prominent at a scattering angle of
$3^\circ$. The observed 
resonance coincides with the spin-dipole strength distribution 
obtained in a multipole analysis of $^{58}$Ni$(n,p)^{58}$Co 
cross sections \cite{el-kateb94}.
\section{Acknowledgements}
This work was supported by the European Union through the Human
Capital and Mobility Program under contract number ERB4050PL932447 
and the Large-Scale Facility program
LIFE under contract number ERBFMGECT980125 and the Land Nordrhein-Westfalen. 
It was performed as part of the research program of the Fund for 
Scientific Research (FSR) Flanders and the Stichting voor Fundamenteel
Onderzoek der Materie (FOM) with financial support from
the Nederlandse Organisatie voor Wetenschappelijk Onderzoek (NWO). \\
\par
\small
$^{**}$\underline{The EUROSUPERNOVA collaboration} \\
R. Bassini$^{f)}$, 
C. B\"aumer$^{a)}$, 
F. Bauwens$^{b)}$,
A.M. van den Berg$^{g)}$,
N. Blasi$^{f)}$,
C. De Coster$^{b)}$,
A.E.L. Dieperink$^{g)}$,
F. Ellinghaus$^{a)}$,
D. Frekers$^{a)}$,
D. De Frenne$^{b)}$,
M. Hagemann$^{b)}$,
V.M. Hannen$^{g)}$,
M.N. Harakeh$^{g)}$,
R. Henderson$^{i)}$,
K. Heyde$^{b)}$,
J. Heyse$^{b)}$,
M.A. de Huu$^{g)}$,
E. Jacobs$^{b)}$,
B.A.M. Kr\"usemann$^{g)}$,
K. Langanke$^{h)}$,
R. De Leo$^{e)}$,
M. Malatesta$^{f)}$,
S. Micheletti$^{f)}$,
M. Mielke$^{a)}$,
P. von Neumann-Cosel$^{c)}$,
M. Pignanelli$^{f)}$,
S. Rakers$^{a)}$,
B. Reitz$^{c)}$,
A. Richter$^{c)}$,
R. Schmidt$^{a)}$,
G. Schrieder$^{c)}$,
H. Sohlbach$^{d)}$,
A. Stascheck$^{c)}$,
S.Y. van der Werf$^{g)}$,
H. De Witte$^{b)}$,
H.J. W\"ortche$^{a)}$ \\
a) Westf\"alische Wilhelms-Universit\"at, M\"unster, Germany \\
b) Universiteit Gent, Gent,Belgium \\
c) TU Darmstadt, Darmstadt, Germany\\
d) M\"arkische Fachhochschule, Iserlohn,Germany\\
e) Universita degli Studi Bari, Bari, Italy \\
f) INFN, Milano, Italy \\
g) Kernfysisch Versneller Instituut, Groningen, The Netherlands \\
h) Aarhus Universitet, Aarhus, Denmark \\
i) TRIUMF, Vancouver, Canada \\
\normalsize
\bibliographystyle{elsart-num}
\bibliography{snova}
\end{document}